\documentstyle[prb,aps,twocolumn,eqsecnum]{revtex}
\begin{document}
\draft
\title{Exact Fermi-edge singularity exponent  in a Luttinger liquid}
\author{Andrei Komnik and Reinhold Egger} 
\address{Fakult\"at f\"ur Physik, Albert-Ludwigs-Universit\"at, 
Hermann-Herder-Stra{\ss}e 3, D-79104 Freiburg, Germany}
\author{Alexander O. Gogolin}
\address{Department of Mathematics, Imperial College, 180 Queen's Gate,
London SW7 2BZ, United Kingdom}
\date{Date: \today}
\maketitle
\begin{abstract}
We report the exact calculation of  the  
Fermi-edge singularity exponent for correlated electrons 
in one dimension (Luttinger liquid).
Focusing on the special interaction parameter $g=1/2$,
the asymptotic long-time behavior can be obtained using the
Wiener-Hopf method. The result confirms the previous
assumption of an open boundary fixed point.
In addition, a dynamic $k$-channel Kondo impurity is studied 
via Abelian bosonization for $k=2$ and $k=4$.
It is shown that the corresponding orthogonality exponents 
are related to the orthogonality
exponent in a Luttinger liquid.
\end{abstract}

\narrowtext

\section{Introduction}

The singular response of conduction electrons in metals
to a transient potential is a centerpiece of 
condensed-matter theory.  It is related 
to basic phenomena such as Anderson's orthogonality 
catastrophe,\cite{anderson67}
the Fermi-edge singularity in x-ray absorption spectra,
\cite{nozieres69,schotte69,mahan90}
 and the Kondo effect.\cite{hewson93} While the
response of conduction electrons  to a time-dependent perturbation
is well understood by now for the case of uncorrelated conduction electrons,
it is both of fundamental and practical  interest to gain insight
into the behavior of correlated electrons under transient potentials.
The Luttinger liquid model for correlated one-dimensional (1D) electrons 
\cite{luttinger,haldane81,schulz95} provides
the perhaps simplest case study for such an investigation. 
The correlation degree is measured by a single
dimensionless interaction strength $g$; for 
repulsive Coulomb interactions, one has $g<1$.
Due to recent fabrication advances, experimental
applications for the Fermi-edge singularity in a
Luttinger liquid have emerged in the 1D quantum
wires in semiconductor heterostructures.\cite{exp,tarucha95,yacoby96}  

The Fermi-edge singularity reflects the instability of the electron sea
to the local core-hole potential generated by an x-ray 
absorption process.\cite{nozieres69,schotte69,mahan90}
Quantities of particular interest are, e.g., the overlap integral between
the ground-state wavefunctions $\langle f |
i \rangle$ with and without the perturbation, 
or the core-hole Green's function $D(t)=\langle i|
d^\dagger(t) d(0) | i \rangle$, which is connected to
the x-ray photoemission spectrum. 
Here, the operator $d$ empties the core-hole level. 
As discussed below, the overlap 
integral and the core-hole Green's functions are closely
related quantities, and both are characterized by  the
same universal exponent $\alpha$.  
This exponent enters the x-ray absorption rate for frequencies
very close to the threshold frequency. 
The exponents of other response functions of interest \cite{mahan90}
can be related to $\alpha$ and will not be discussed in what follows.

There have been several previous works related to the
Fermi-edge singularity and the orthogonality catastrophe in a Luttinger
 liquid.\cite{lee92,ogawa92,gogolin93,kane94,prokofev94,affleck94,oreg96,qin96}
The forward-scattering contribution due to a transient potential
was evaluated in Refs.\onlinecite{lee92} and \onlinecite{ogawa92},
and since the absorption rate factorizes into  forward- and 
backward-scattering parts,\cite{gogolin93}
 we here focus on the more interesting
backscattering contribution solely.
Essentially based on the assumption of an open boundary fixed point,
\cite{fabrizio95} the  Fermi-edge 
singularity exponent $\alpha=1/8$ has been
found for all $g<1$ by various techniques in 
Refs.\onlinecite{gogolin93,kane94,prokofev94,affleck94}.
The assumption of an
open boundary fixed point has recently been questioned by Oreg and
Finkel'stein \cite{oreg96} who find the result $\alpha=0$ instead. This
discrepancy has also 
been addressed in a recent numerical work.\cite{qin96}
Here we decide the issue by
presenting an exact derivation of the exponent $\alpha=1/8$.

Our calculation is performed at the special  interaction strength $g=1/2$.
This suffices to resolve the controversy about the open boundary
fixed point.  At the special point $g=1/2$,
 the bosonized Luttinger liquid Hamiltonian can be refermionized
in a simple way by employing a suitable chiral fermion basis.
This trick has been exploited previously for conductance calculations
in the presence of a static 
impurity.\cite{fabrizio95,guinea85,weiss88,matveev95}
Due to the transient nature of the core-hole potential, the situation
is more intricate here, but nevertheless it allows for an exact solution
by applying the Wiener-Hopf technique.
We mention that the point $g=1/2$  can allow for exact results
even when the quantity of interest cannot be written in terms of chiral
fermions, see, e.g., a recent calculation of the Friedel oscillation around a
static impurity.\cite{saleur96}

In Ref.\onlinecite{affleck94}, which addresses the
Fermi-edge singularity problem from
the boundary conformal field theory point of view,
it was noted that even in conventional metals 
the final state of the system may have a non-Fermi-liquid
character (albeit the initial state is a Fermi liquid).
This can happen, for instance, when the core hole created in the
absorption process has internal degrees of freedom like spin or
orbital quantum numbers. 
For a final-state impurity leading to a multi-channel Kondo problem,
the x-ray edge exponents were found in Ref.\onlinecite{affleck94}.
On the other hand, considerable progress in the theory of
the $k$-channel Kondo problem has recently been achieved.
In the framework of Abelian bosonization, Toulouse-limit 
type solutions \cite{toulouse69}
were found for the $k=2$ and $k=4$ models.\cite{emery92,fg95}
In the latter two cases, the Toulouse limit turns out to be
equivalent to an impurity problem in a Luttinger liquid.
Confusingly, the x-ray edge exponents of Ref.\onlinecite{affleck94}
for the Kondo problem are different from the $\alpha=1/8$
result for the Luttinger liquid.  This puzzle  is
addressed and resolved here.

The outline of this paper is as follows. In Sec.~\ref{sec:ii},
we briefly discuss  the Luttinger liquid  and 
fermionize the theory at $g=1/2$.  Sec.~\ref{sec:iii} presents
the derivation of an integral equation for the core-hole
Green's function. The  asymptotically exact 
solution of this integral equation
is discussed in Sec.~\ref{sec:iv}. Thereby the 
Fermi-edge singularity exponent $\alpha=1/8$ is found.
In Sec.~\ref{sec:v}, we re-derive the exponents 
for the $k$-channel Kondo problem and show that they are in
fact consistent with the Luttinger liquid result.
Finally, some conclusions are offered in Sec.~\ref{concl}.

\section{Luttinger liquid} \label{sec:ii}

We treat the 1D interacting electron liquid in the framework of
the Luttinger liquid model,\cite{haldane81,schulz95} 
focusing on the spinless case.  The most appealing
theoretical description is offered by the bosonization
treatment, which equivalently expresses the interacting
fermionic system as a harmonic fluid composed of 
bosonic plasmon modes.
In terms of  right- and left-moving 
boson fields $\varphi_{L,R}$ satisfying the algebra $(p=R,L=\pm)$
\begin{equation}\label{a1}
[\varphi_p(x) , \varphi_{p'} (x')] = -i\pi p\, \delta_{p,p'}
\,{\rm sgn}(x-x') \;,
\end{equation}
the Hamiltonian takes the form (we put $\hbar=1$ and
the sound velocity $v=1$)
\begin{equation}\label{hii1}
H_i = \frac{1}{8\pi}\int dx\, [(\partial_x\varphi_R)^2 + 
(\partial_x \varphi_L)^2 ]\;.
\end{equation}
This initial Hamiltonian 
describes the unperturbed Luttinger liquid prior to 
the absorption process at time $t_i=0$, i.e., for 
$t<t_i$.  The local backscattering due to
the core-hole potential acting at $x=0$ and $t>t_i$ 
leads to
\begin{equation}\label{hff1}
H_f= H_i + V  \cos [\sqrt{g}\,(\varphi_R(0)-\varphi_L(0))]\;.
\end{equation}
The spinless Luttinger liquid is characterized by the dimensionless
interaction strength parameter $g$, where
 $g=1$ is the non-interacting Fermi liquid value, and $g<1$ 
corresponds to repulsive  Coulomb interactions.
The strength of the transient backscattering is denoted by $V$.

In what follows, it is advantageous to introduce chiral 
right-moving boson fields,
\[
\phi_\pm(x) = [\varphi_R(x)\mp \varphi_L(-x)]/\sqrt{2}\;,
\]
which from Eq.~(\ref{a1}) obey the algebra
\begin{equation} \label{a2}
[\phi_p(x) , \phi_{p'} (x')] = -i\pi\,  \delta_{p,p'} \,{\rm sgn}(x-x') \;.
\end{equation}
Written in terms of $\phi_\pm(x)$, 
the Hamiltonian decouples into a sum of odd and even fields,
\begin{eqnarray} \nonumber
H_i &=& \frac{1}{8\pi} \int dx \,  [(\partial_x\phi_+)^2 + 
(\partial_x \phi_-)^2 ]  \;,\\  \label{xxxx}
H_f &=& H_i + V \cos[\sqrt{2g}\, \phi_+(0) ] \;.
\end{eqnarray}
Because $\phi_-$ does not couple to the impurity and at the same
time commutes with $\phi_+$, see Eq.~(\ref{a2}), we drop
the $(\partial_x \phi_-)^2$ term in $H_i$ in the following. 

Let us now consider the special value $g=1/2$,
where the Hamiltonian $H_f$ can be re-fermionized by 
using the chiral right-moving fermion field operator
\begin{equation}\label{bos1}
\psi (x) = (2\pi a)^{-1/2} \,\exp[i \phi_+(x)]\;,
\end{equation}
with  a short-distance cutoff $a$ (say, a lattice spacing).
Absorbing the factor $\sqrt{2\pi a}$ into the backscattering
strength $V$, we obtain the fermionic description
\begin{eqnarray*}
H_i &=& -  i\int dx\,\psi^\dagger(x)  \partial_x \psi (x) \;, \\
H_f &=& H_i+  \frac{V}{2} \,[\psi(0)+\psi^\dagger(0)] \;.
\end{eqnarray*}
To proceed further, it is convenient to use a trick introduced
by Matveev\cite{matveev95} and write
\begin{equation}
\psi(x)= (c+c^\dagger)\, \tilde{\psi}(x) \;,
\end{equation}
with new fermion operators $c$ and $\tilde{\psi}(x)$. 
It is a simple matter to check that $\psi$ is still fermionic
in this representation. Since $(c+c^\dagger)^2=1$,
 the Hamiltonian takes the form
\begin{eqnarray*}
H_i &=& - i\int dx\,\tilde{\psi}^\dagger(x)  \partial_x
\tilde{\psi} (x)\;,  \\
H_f &=& H_i+  \frac{V}{2} (c+c^\dagger) [\tilde{\psi}(0) - 
\tilde{\psi}^\dagger(0)] \;.
\end{eqnarray*}

Finally, we switch to
a representation in terms of real Majorana fermion operators, 
\begin{eqnarray} \label{majorana}
c = (b+ia)/\sqrt{2} \;, && \quad c^\dagger=(b-ia)/\sqrt{2}\;, \\
\tilde{\psi}(x) = (\eta(x)+i\xi(x))/\sqrt{2} \;, && \quad
\tilde{\psi}^\dagger(x) = (\eta(x)-i\xi(x))/\sqrt{2} \;. 
\nonumber
\end{eqnarray}
Here, we have the anticommutator algebra
\begin{eqnarray*}
[\xi(x),\xi(x')]_+ & = & [\eta(x),\eta(x')]_+ = \delta(x-x') \;, \\
 \, [ \eta(x),\xi(x') ]_+  &=& 0 \;,
\end{eqnarray*}
and likewise for $a,b$.   Written in terms of these new fields,
\begin{eqnarray}
H_i &=& -(i/2) \int dx\, [\eta(x)  \partial_x\eta(x) + 
\xi(x) \partial_x \xi(x) ]\;, \nonumber \\
H_f &=& H_i+  iV b \xi(0) \label{ferm2}\;.
\end{eqnarray}
Since the $\eta$ field does not couple
to the local backscattering potential, we  drop it in the following. 
These expressions are the basis of the subsequent treatment.

\section{Integral equation for the core-hole Green's function}
\label{sec:iii}

One relevant dynamical quantity which allows extraction of the
exponent $\alpha$ is the core-hole Green's function. It can be
defined as
\begin{equation}\label{ddef}
D(t) = \left\langle e^{iH_i t} e^{-iH_f t} \right \rangle^{}_0 \;,
\end{equation}
where $\langle \cdots \rangle^{}_0$ denotes a ground-state
average using 
\begin{equation}
H_i= -(i/2) \int dx \, \xi(x) \partial_x \xi(x) \;.
\end{equation}
 At long times, the core-hole Green's function behaves like
\begin{equation}\label{general}
D(t) \sim e^{-i\Delta E \,t}\, t^{-\alpha}\;,
\end{equation}
with the energy shift $\Delta E$ of the 
core-hole level and the Fermi-edge singularity 
 exponent  $\alpha$. The latter quantity is the aim of our calculation.

For $t\to \infty$, the function (\ref{ddef}) gives the
{\em overlap} of the initial and the final many-body
wavefunction, which vanishes with 
the dimensionless system size $N$ as 
\begin{equation}\label{dinf}
D(\infty) \sim N^{-\alpha/2} \;.
\end{equation}
This is the famous Anderson orthogonality catastrophe.\cite{anderson67}
The factor $1/2$ in the exponent compared with
Eq.~(\ref{general}) is explained as follows.
For the
overlap (\ref{dinf}), the backscattering potential is switched on at
time $t_i=0$ but never switched off. However, at finite times $t$,
for the core-hole Green's function $D(t)$, the potential
is effectively switched off at time $t$. Therefore one has two
switching processes which  results in a doubling of the exponent
compared to the overlap integral.\cite{gogolin93}

Using Eqs.~(\ref{ferm2}) and (\ref{ddef}), and switching to
imaginary time $\tau=it$, the core-hole
Green's function becomes
\begin{equation}
D(\tau_i,\tau_f) = \left\langle 
{\cal T} \exp\left[-iV\int_{\tau_i}^{\tau_f}
d\tau\, b(\tau) \xi(\tau) \right] \right \rangle^{}_0 \;,
\end{equation}
where $\xi(\tau) \equiv \xi(x=0,\tau)$ and ${\cal T}$ is the 
time-ordering operator. This can be written more suitably 
by first introducing the modified function
\begin{equation}
D_\lambda(\tau_i,\tau_f) = \left\langle 
{\cal T} \exp\left[-i\lambda V\int_{\tau_i}^{\tau_f}
d\tau\, b(\tau) \xi(\tau) \right] \right \rangle^{}_0 \;,
\end{equation}
which fulfills (time integrations always go from $\tau_i$
to $\tau_f$ in what follows)
\begin{equation}\label{ppp}
\frac{\partial}{\partial\lambda} D_\lambda = - iV \left
[ \int d\tau\,
Y_\lambda(\tau,\tau) \right ]  D_\lambda
\end{equation}
with the auxiliary function
\begin{equation}\label{ydef}
Y_\lambda(\tau,\tau') = \langle {\cal T} b(\tau) \xi(\tau') 
\rangle^{}_\lambda\;.
\end{equation}
Here, $\langle \cdots \rangle^{}_\lambda$ refers to a ground-state
 average using
\begin{equation}\label{hlambda}
 H_\lambda = H_i + i\lambda V b \xi(0)\;.
\end{equation}
Note that since the backscattering is only present for times
$\tau$ in the interval
$\tau_i < \tau < \tau_f$, the quantity (\ref{ydef}) 
depends on $\tau_i, \tau_f$.
Substituting $\lambda \to \lambda V$, we then obtain from
Eq.~(\ref{ppp})
\begin{equation}\label{ddd}
\ln D (\tau_i, \tau_f) = - i \int_0^V d\lambda \int d\tau \,Y_\lambda 
(\tau,\tau)\;.
\end{equation}
To exploit this equation, we will now formulate and solve 
an integral equation for the function $Y_\lambda$.

For that purpose, we first need the free Green's functions
\begin{eqnarray*}
G_b(\tau-\tau') &\equiv& \langle {\cal T} b(\tau) b(\tau') \rangle^{}_0
\;,\\
 G_\xi(\tau-\tau') &\equiv& \langle {\cal T} \xi(\tau) \xi(\tau') 
\rangle^{}_0 \;.
\end{eqnarray*}
They can  easily be computed from their respective equation
of motion.  Since $\partial_\tau b = [b, H_i] = 0$ and
$b^2 = (c+c^\dagger)^2/2 = 1/2$ [see Eq.~(\ref{majorana})], this
reads 
\[
\frac{\partial}{\partial\tau} G_b(\tau-\tau') = \delta(\tau-\tau') \;,
\]
with the solution
\begin{equation} \label{ggb}
G_b(\tau-\tau') = (1/2) \, {\rm sgn}(\tau-\tau')\;.
\end{equation}
For $G_\xi(\tau-\tau')$, we first consider the correlation
function $\langle {\cal T} \xi(x,\tau) \xi(x',\tau') \rangle^{}_0$.
With $\partial_\tau \xi(x) = [\xi(x), H_i]  =  i \partial_x \xi(x)$,
its  equation of motion reads
\[
\left(\frac{\partial}{\partial\tau} - i \frac{\partial}{\partial x}
 \right) \langle {\cal T} \xi(x,\tau) \xi(x',\tau') \rangle_0^{}
= \delta(\tau-\tau') \delta(x-x')\;.
\]
Specializing the solution to $x=x'=0$, we obtain
\begin{equation}\label{gxi}
G_\xi(\tau-\tau') = \frac{1}{2\pi (\tau-\tau')}\;.
\end{equation}

The equation-of-motion method can similarly be employed
to derive an integral equation for the function $Y_\lambda(\tau,\tau')$
defined in Eq.~(\ref{ydef}). Under
$H_\lambda$ [note that the factor
 $V$ in Eq.~(\ref{hlambda}) has meanwhile been absorbed
in $\lambda$], the time evolution of $b$ and
$\xi(x)$ is governed by
\begin{eqnarray*}
\frac{\partial}{\partial\tau} b &=& [b, H_\lambda] = i\lambda\xi(0) \\
\frac{\partial}{\partial\tau} 
\xi(x) &=& [\xi(x), H_\lambda] = i\frac{\partial}{\partial x}
 \xi(x) -i\lambda\delta(x) b\;.
\end{eqnarray*}
Applying a time derivative to Eq.~(\ref{ydef}), we have
\begin{equation} \label{fp1}
\frac{\partial}{\partial\tau} Y_\lambda(\tau,\tau') = i\lambda 
\langle {\cal T} \xi(\tau) \xi(\tau') \rangle^{}_\lambda \;.
\end{equation}
Similarly, applying a time derivative to the function
$\langle {\cal T} \xi(x,\tau) \xi(\tau') \rangle^{}_\lambda$
generates the equation of motion
\begin{eqnarray*}
&& \left ( \frac{\partial}{\partial\tau} - i \frac{\partial}{\partial x}\right)
 \langle {\cal T} \xi(x,\tau)
 \xi(\tau') \rangle^{}_\lambda = \\ &&\quad 
 \delta(x) \delta(\tau-\tau') - i\lambda \delta(x)
Y_\lambda(\tau, \tau')\;.
\end{eqnarray*}
Employing the free Green's function (\ref{gxi}), the
solution of this equation (for $x=0$) is
\begin{equation}\label{fp2}
\langle {\cal T} \xi(\tau) \xi(\tau') \rangle^{}_\lambda
= G_\xi(\tau-\tau') - i\lambda\int d\tau_1
\, G_\xi(\tau-\tau_1) Y_\lambda(\tau_1,\tau')\;.
\end{equation}
On the other hand, from Eqs.~(\ref{ggb}) and (\ref{fp1}), we have
\begin{equation}\label{fff}
Y_\lambda(\tau,\tau') = i\lambda \int d\tau_1 \,
G_b(\tau-\tau_1) \langle {\cal T} \xi(\tau_1)
 \xi(\tau') \rangle^{}_\lambda\;.
\end{equation}
Inserting Eq.~(\ref{fp2}) into Eq.~(\ref{fff}) results in the {\em integral
equation} 
\begin{equation}
Y_\lambda(\tau,\tau') = -i \lambda  f(\tau,\tau') - 
\lambda^2 \int d\tau_1 \, f(\tau,\tau_1) Y_\lambda(\tau_1, \tau') \;.
\end{equation}
The {\em singular kernel} $f(\tau,\tau')$ is defined by
\begin{eqnarray} \label{kernel}
f(\tau,\tau') &=& \int d\tau_1 \,G_b(\tau-\tau_1) G_\xi(\tau'-\tau_1)\\
\nonumber
&=& \frac{1}{4\pi} \int d\tau_1 \,
\frac{{\rm sgn}(\tau-\tau_1)}{\tau'-\tau_1} \;.
\end{eqnarray}

Let us conclude this section by summarizing and slightly simplifying
the main result. Redefining $Y_\lambda \to Y_\lambda/\lambda$, 
we have derived the integral equation 
\begin{equation}\label{inteq}
Y_\lambda(\tau,\tau') = -i   f(\tau,\tau') - 
\lambda^2 \int d\tau_1 \, f(\tau,\tau_1) Y_\lambda(\tau_1, \tau') 
\end{equation}
with the kernel $f(\tau,\tau')$ given in  Eq.~(\ref{kernel}). 
Once this equation is solved, the core-hole Green's function and
the overlap integral can be determined from 
\begin{equation}\label{dddd}
\ln D (\tau_i, \tau_f) = - i \int_0^V d\lambda \lambda
\int_{\tau_i}^{\tau_f} d\tau \,Y_\lambda 
(\tau,\tau)\;,
\end{equation}
see Eq.~(\ref{ddd}).
The solution of the  integral equation (\ref{inteq})
for the overlap integral, i.e., for $\tau_i=0$ and $\tau_f\to \infty$,
 is discussed in the next section.

As a simple consistency 
check for these expressions, the short-time  behavior of
the core-hole Green's function can be calculated by 
perturbation theory in $\lambda$ from Eqs.~(\ref{inteq}) 
and (\ref{dddd}). The result coincides with the 
result from Ref.\onlinecite{gogolin93}
obtained by straightforward perturbation theory
in $V$ using the original bosonized picture.

\section{Wiener-Hopf solution for the overlap integral} \label{sec:iv}

We now turn to the calculation of the overlap integral
($\tau_i=0, \tau_f \to \infty$). Then the
integral equation (\ref{inteq}) can be solved using the
Wiener-Hopf method.\cite{mus53,hamann71} For
$ \tau, \tau'  > 0$, the singular kernel (\ref{kernel}) takes the form
\begin{equation}
f(\tau,\tau') = f_1(\tau-\tau') + f_2(\tau') \;,
\end{equation}
where
\begin{eqnarray*}
f_1(\tau-\tau') &=& -\frac{1}{2\pi} \,\ln|\tau-\tau'| \\
f_2(\tau') &=& \frac{1}{4\pi}\, [\ln(\tau_f-\tau')+ \ln (\tau')] \;.
\end{eqnarray*}
The function $f_2(\tau')$ requires a finite cutoff $\tau_f$.
However, it will turn out that terms related to $f_2$ do not contribute to
the overlap integral, and therefore this formal divergence is not 
detrimental to us.

We next switch to (partial) Fourier transforms,
\begin{eqnarray*}
Y_\lambda(\tau,\tau') &=& \int \frac{d\omega}{2\pi}
e^{-i\omega \tau} \tilde{Y}_\lambda(\omega,\tau')\\
f_1(\tau-\tau') &=& \int \frac{d\omega}{2\pi} 
e^{-i\omega(\tau-\tau')} \tilde{f}_1(\omega) \;,
\end{eqnarray*}
where
\begin{equation}\label{f1}
\tilde{f}_1(\omega) = \frac{1}{2|\omega|} \;.
\end{equation}
For application of the Wiener-Hopf method, it is necessary to
introduce the functions ($\epsilon\to 0$)
\begin{equation}
\tilde{Y}^{(\pm)}_\lambda (\omega,\tau') \equiv
\pm \int \frac{d\omega'}{2\pi i} \frac{\tilde{Y}_\lambda(\omega',
\tau')}{\omega'-\omega\mp i \epsilon} \;,
\end{equation}
which are analytic functions in the upper/lower $\omega$-plane
$\Pi^{(\pm)}$, respectively.  Naturally, the sum of both
reproduces the function itself, $\tilde{Y}_\lambda=
\tilde{Y}_\lambda^{(+)} + \tilde{Y}_\lambda^{(-)}$.

With these conventions, the integral equation (\ref{inteq}) is
equivalently  expressed in the form
\begin{eqnarray} \label{inteq2}
\tilde{Y}_\lambda(\omega,\tau') &=& 
-i e^{i\omega \tau'} \tilde{f}_1(\omega) - 2\pi i \delta(\omega) f_2(\tau') \\
&-& \nonumber
 \lambda^2 \tilde{f}_1(\omega) \tilde{Y}_\lambda^{(+)} (\omega,\tau')
\\ &-& 2\pi\lambda^2 \delta(\omega) \int d\tau_1 f_2(\tau_1) 
Y_\lambda(\tau_1,\tau') \nonumber \;.
\end{eqnarray}
This integral equation is of  Wiener-Hopf form. That is made
apparent by introducing the quantities $X_\lambda^{(\pm)}(\omega)$
fulfilling
\begin{equation} \label{xx}
X_\lambda^{(+)} (\omega) X_\lambda^{(-)}(\omega) = 
\frac{1}{1+\lambda^2 \tilde{f}_1(\omega) } \;,
\end{equation}
with the explicit form
\begin{equation} \label{xpm}
\ln  X_\lambda^{(\pm)} (\omega) =\mp
\int\frac{d\omega'}{2\pi i} 
\frac{\ln(1+\lambda^2 \tilde{f}_1(\omega'))}{\omega'-\omega\mp
i\epsilon}\;.
\end{equation}
With these definitions, Eq.~(\ref{inteq2}) becomes
\begin{eqnarray} \label{inteq3}
&& \frac{\tilde{Y}^{(+)}_\lambda(\omega,\tau')}{X_\lambda^{(+)}(\omega)}
+ \tilde{Y}^{(-)}_\lambda(\omega,\tau') X_\lambda^{(-)}(\omega) = \\
&& \quad -i e^{i\omega \tau'} \tilde{f}_1(\omega) X_\lambda^{(-)}(\omega)
\nonumber
 - 2\pi i \delta(\omega) f_2(\tau') X_\lambda^{(-)}(\omega) \\
&& \quad - 2\pi\lambda^2 \delta(\omega) X_\lambda^{(-)}(\omega)
\int d\tau_1 f_2(\tau_1) 
Y_\lambda(\tau_1,\tau') \nonumber\;.
\end{eqnarray}
As mentioned above, all terms related to $f_2(\tau')$ do not
contribute to the solution of Eq.~(\ref{inteq3}). This can
be seen by inspecting the $\omega\to 0$ limit of 
$X_\lambda^{(-)}(\omega)$. From Eqs.~(\ref{f1}) and 
(\ref{xpm}), we have
\begin{eqnarray*}
\ln X_\lambda^{(-)}(\omega=0) &=& \int 
\frac{d\omega'}{2\pi i} 
\frac{\ln(1+\frac{\lambda^2}{2|\omega'|})}{\omega'+i\epsilon}\\
&=& - \int_0^\infty d\omega'\, \frac{\epsilon/\pi}{\omega^{\prime 2}+
\epsilon^2} \ln(1+\frac{\lambda^2}{2\omega'}) \;.
\end{eqnarray*}
As $\epsilon\to 0$, this expression approaches $-\infty$ such that
\[
 \lim_{\omega\to 0} X_\lambda^{(-)}(\omega)= 0\;.
\]
Therefore, when integrating over $\omega$ in Eq.~(\ref{inteq3}),
all terms containing  $f_2(\tau')$ drop out due to the $\delta(\omega)$
factors.

In the end, only the translationally invariant part  $f_1$ of the
kernel matters,
\begin{equation}\label{inteq4}
\tilde{Y}_\lambda^{(+)}(\omega,\tau') = 
\frac{-i \tilde{f}_1(\omega) e^{i\omega\tau'}}
{1+\lambda^2 \tilde{f}_1(\omega)} 
- X_\lambda^{(+)}(\omega) X_\lambda^{(-)}(\omega) 
\tilde{Y}_\lambda^{(-)}(\omega,\tau') \;.
\end{equation}
Since Eq.~(\ref{dddd}) can be written in the form
\begin{equation}\label{ddddd}
\ln D (\tau_f)= -i \int_0^V d\lambda \lambda \int_0^{\tau_f} d\tau
 \int \frac{d\omega}{2\pi}
e^{-i\omega\tau} \tilde{Y}_\lambda^{(+)}(\omega,\tau)\;,
\end{equation}
we obtain two contributions to $D(\tau_f)$ from Eq.~(\ref{inteq4}).

The first term generates the energy shift $\Delta E$ of the core-hole
level in Eq.~(\ref{general}),
\[
\ln D^{(1)} (\tau_f) = - i \Delta E (-i\tau_f) 
\]
where
\begin{equation}\label{shift}
\Delta E= \int_0^V d\lambda \lambda \int \frac{d\omega}{2\pi}
\frac{\tilde{f}_1(\omega)}{1+\lambda^2 \tilde{f}_1(\omega)} \;.
\end{equation}
After analytic continuation to real time, $-i \tau_f \to t$, this 
gives exactly the $\exp(-i\Delta E \,t)$ factor in Eq.~(\ref{general}).
The value of $\Delta E$ can be computed from Eq.~(\ref{shift}),
but it is not of immediate interest to us in the following.

The second term in Eq.~(\ref{inteq4}) is responsible for the
{\em orthogonality catastrophe}.  For the overlap integral, this
gives  
\begin{eqnarray} \label{d1}
\ln D (\infty) &=& i \int_0^V d\lambda \lambda\int_0^\infty
 d\tau e^{-\epsilon \tau} \int \frac{d\omega}{2\pi}
e^{-i\omega\tau} \\ \nonumber
&\times& X_\lambda^{(+)}(\omega) X_\lambda^{(-)}(\omega)
\tilde{Y}_\lambda^{(-)}(\omega,\tau) \;.
\end{eqnarray}
To get $\tilde{Y}_\lambda^{(-)}(\omega,\tau)$, we
 now apply the Wiener-Hopf trick to Eq.~(\ref{inteq3})
[note that the $f_2$ terms can be omitted]: 
The r.h.s.~of Eq.~(\ref{inteq3}) can be separated into two
functions which are analytic in $\Pi^{(\pm)}$, respectively.
This yields
\begin{equation}
\tilde{Y}_\lambda^{(-)}(\omega,\tau) = 
\frac{i}{X_\lambda^{(-)}(\omega)}
\int \frac{d\omega' }{2\pi i}
\frac{e^{i\omega' \tau}\tilde{f}_1(\omega') X_\lambda^{(-)}(\omega')}
{\omega'-\omega +i\epsilon}\;,
\end{equation}
which is now inserted into Eq.~(\ref{d1}) and gives
\begin{eqnarray} \label{d2}
\ln D (\infty) &=& -\int_0^V d\lambda \lambda\int_0^\infty
 d\tau e^{-\epsilon \tau} \int \frac{d\omega}{2\pi}
\int \frac{d\omega'}{2\pi i} \\ &\times& \nonumber
e^{-i\omega\tau}  X_\lambda^{(+)}(\omega)
 \; \frac{e^{i\omega' \tau}\tilde{f}_1(\omega') X_\lambda^{(-)}(\omega')}
{\omega'-\omega +i\epsilon}\;.
\end{eqnarray}
We next apply the relation
\[
\lambda^2 \tilde{f}_1(\omega') X_\lambda^{(-)}(\omega')
= \frac{1}{X_\lambda^{(+)}(\omega')} - X_\lambda^{(-)}(\omega')\;,
\]
which follows from Eq.~(\ref{xx}). 
This gives from Eq.~(\ref{d2}) the expression
\begin{eqnarray*}
\ln D (\infty) &=& -\int_0^V \frac{d\lambda}{\lambda}\int_0^\infty
 d\tau e^{-\epsilon \tau} \int \frac{d\omega}{2\pi}
\int \frac{d\omega'}{2\pi i} e^{-i(\omega-\omega')\tau}  \\ &\times& 
\left(\frac{1}{X_\lambda^{(+)}(\omega')} - X_\lambda^{(-)}(\omega')
\right) \frac{X_\lambda^{(+)}(\omega)} {\omega'-\omega +i\epsilon}\;.
\end{eqnarray*}
The $1/X_\lambda^{(+)}(\omega')$ term does not contribute,
such that we obtain with the variables $\omega_1=\omega'/\lambda^2$,
$\omega_2=\omega/\lambda^2, \gamma= \lambda^2$, and
$\epsilon \to \epsilon/\lambda^2$
\begin{eqnarray} \label{d4}
\ln D (\infty) &=& \frac12 \int_0^{V^2} d\gamma\int_0^\infty
 d\tau e^{-\epsilon\gamma \tau} \int \frac{d\omega_1}{2\pi}
\int \frac{d\omega_2}{2\pi i} \\ &\times& \nonumber
e^{i\gamma\tau(\omega_1-\omega_2)}  
\frac{X_1^{(+)}(\omega_2) X_1^{(-)}(\omega_1)}
{\omega_1-\omega_2 +i\epsilon}\;.
\end{eqnarray}
Remarkably, the integrand is symmetric under the
exchange  of $\tau$ and $\gamma$, since only their
product appears.
In Eq.~(\ref{d4}) we have also exploited the fact
\[
X_\lambda^{(\pm)} (\omega)  = X_1^{(\pm)} (\omega/\lambda^2) \;,
\]
which is apparent from the definition (\ref{xpm}).

Naturally, Eq.~(\ref{d4}) is an infra-red divergent expression since
the overlap vanishes for the infinite system. For a 
finite system of dimensionless length $N$, the shortest possible timescales 
are of the order $\sim 1/N$, and the $\tau$ integration 
in Eq.~(\ref{d4}) should be cut off from below at a scale 
$\tau_0/N$ (where $\tau_0$ is a timescale of the infinite system).
Because of the formal symmetry of the integrand of Eq.~(\ref{d4})
under exchange of $\gamma$ and $\tau$, we can alternatively
cut off the $\gamma$ integration from below.  The
$\tau$ integration can then be carried out and gives
a factor $i/[\gamma(\omega_1-\omega_2+i\epsilon)]$.
In the end, the result becomes
\begin{equation} \label{d5}
\ln D (\infty) = -\frac12 \int_{\tau_0/N}^{V^2}\frac{d\gamma}{\gamma}
 \int \frac{d\omega_1}{2\pi i} \int \frac{d\omega_2}{2\pi i} 
\frac{X_1^{(+)}(\omega_2) X_1^{(-)}(\omega_1)}
{(\omega_1-\omega_2 +i\epsilon)^2}\;.
\end{equation}
For $N\to\infty$, this yields indeed the power law (\ref{dinf}) for the
overlap integral, and we can read off the exponent $\alpha$ by
comparing Eqs.~(\ref{dinf}) and (\ref{d5}),
\begin{equation} \label{ex1}
\alpha = 
 \int \frac{d\omega_1}{2\pi i} \int \frac{d\omega_2}{2\pi i} 
\frac{X_1^{(+)}(\omega_2) X_1^{(-)}(\omega_1)}
{(\omega_1-\omega_2 +i\epsilon)^2}\;.
\end{equation}

The calculation of the Fermi-edge singularity
exponent is now straightforward.
Carrying out the $\omega_2$ integration  in Eq.~(\ref{ex1}), one obtains
\begin{eqnarray*}
\alpha &=& \int \frac{d\omega}{2\pi i}
X_1^{(-)}(\omega) X_1^{(+)}(\omega) \frac{\partial}{\partial\omega} \ln 
X_1^{(+)}(\omega) \\
&=& - \int \frac{d\omega}{2\pi i}
\frac{1}{1+\frac{1}{2|\omega|}}
 \frac{\partial}{\partial\omega} \int \frac{d\omega'}
{2\pi i} \frac{\ln (1+\frac{1}{2|\omega'|})}{\omega'-\omega-i\epsilon} \;,
\end{eqnarray*}
where we have used Eq.~(\ref{xpm}) in the second line.
After a partial integration and the rescaling $\omega_1=2\omega$,
 $\omega_2=2\omega'$, this takes the form
\begin{eqnarray*}
\alpha &=& -\frac{1}{\pi^2} \int_0^\infty d\omega_1
\int_0^\infty d\omega_2
\frac{\omega_1}{(\omega_2^2-\omega_1^2) (\omega_1+1)
(\omega_2+1)}\\
&=& \frac{1}{2\pi^2} (I_+ +I_-) \;.
\end{eqnarray*}
The quantities $I_\pm$  are
\[
I_\pm = \int_0^\infty d\omega_1
\int_0^\infty d\omega_2 \frac{1}{(\omega_1+1)
(\omega_2+1) (\omega_1\pm\omega_2)} \;,
\]
 with the result $I_+=\pi^2/4$ and $I_-=0$.
In the end, we  arrive at the central result of this calculation,
$\alpha=1/8$. This is the exact Fermi-edge singularity exponent
for a Luttinger liquid.

\section{Orthogonality exponents for the multichannel Kondo case}
\label{sec:v}

In the following, we discuss the relationship between the 
orthogonality exponent $\alpha=1/8$  in
a Luttinger liquid and the respective exponent for a
Kondo impurity interacting with $k$ uncorrelated conduction
electron channels.
If the core hole has internal degrees of freedom, $d\to d_\sigma$,
the final-state Hamiltonian takes the form
\begin{equation} \label{HKfin}
H_f = H_i + I\,\vec{s}\vec{J}(0)\;,
\end{equation}
where $H_i$ describes the uncorrelated conduction electrons 
prior to the x-ray absorption process.
The following consideration is restricted to 
$\sigma$ denoting a (pseudo)spin-1/2 index,
such that the impurity spin operator is 
$\vec{s}=d^\dagger_\sigma\vec{\tau}_{\sigma,\sigma'}
d^{\phantom{\dagger}}_{\sigma'}$ with 
the spin-1/2 matrices $\vec{\tau}$.
The conduction electron spin density operators are defined by
\[
\vec{J}(x)= 
\sum\limits_{i=1}^k \psi^\dagger_{i\sigma}(x)\vec{\tau}_{\sigma,\sigma'}
\psi^{\phantom{\dagger}}_{i\sigma'}(x)
\]
for $k$ conduction electron channels.
Without loss of generality, $x$ can be understood as a 
1D coordinate.\cite{comment} As found in Ref.\onlinecite{affleck94},
the x-ray edge exponent for the problem (\ref{HKfin}) is 
\begin{equation}
\label{alphaK}
\alpha_k = \frac{3}{2(2+k)}\;.
\end{equation}

We bosonize the electron field operators in the standard way,
see Eq.~({\ref{bos1}),
\[
\psi_{i\sigma}(x)=\frac{1}{\sqrt{2\pi a}}\,
\exp[i\phi_{i\sigma}(x)]\;,
\]
where $\phi_{i\sigma}(x)$ is a chiral (right-moving) boson field.
Here only spin degrees of freedom are relevant, i.e.,
only the combinations
\[
\phi_i(x)=\frac{1}{\sqrt{2}}\left[\phi_{i\uparrow}(x)-
\phi_{i\downarrow}(x)\right]
\]
enter the bosonized form of the spin density operators,
\begin{eqnarray} \label{Jpsi}
J^-(x)&=&\sum\limits_{i=1}^k \psi^\dagger_{i\downarrow}(x)
\psi^{\phantom{\dagger}}_{i\uparrow}(x)
=
\frac{1}{2\pi a}\sum\limits_{i=1}^k 
{\rm e}^{i\phi_{i}(x)} \;, \\ \nonumber
J^z(x)&=&\frac{\sqrt{k}}{2\sqrt{2}\,\pi}\partial_x\phi_s(x)\;,
\end{eqnarray}
where the total spin field $\phi_s$ is defined as
\[
\phi_s(x)=\frac{1}{\sqrt{k}}\, \sum\limits_{i=1}^k \phi_{i}(x)\;.
\]
The spin density operators $J^a(x)$ ($a=x,y,z$) 
are also referred to as spin currents since they
satisfy the commutation relations of the
$SU(2)$ level-$k$ Kac-Moody current algebra 
\begin{equation} \label{KM}
\left[ J^a(x), J^b(y)\right]=
i\epsilon^{abc}\delta(x-y)J^c(x)+
\frac{ik}{4\pi}\delta'(x-y)\delta^{ab}\;.
\end{equation}

In order to investigate how the result (\ref{alphaK}) fits into
the Abelian treatment of the Kondo problem 
\cite{emery92,fg95} and whether it is consistent 
with the Luttinger liquid exponent $\alpha=1/8$, we next
exploit the following observation.\cite{schotte69}
Assume that we have constructed a unitary transformation $U$
relating the initial and the final Hamiltonian,
\[
U^\dagger H_f U =H_i\;.
\]
Then the core-hole Green's function $D(t)$ can be
represented as
\[
D(t)=\langle i| U^\dagger(t)U(0) |i \rangle \;,
\]
where the time evolution of the unitary operator $U(t)$ is
governed by the initial Hamiltonian,
$U(t)= e^{iH_i t} U  e^{-iH_i t}$.

In general, the Toulouse limit solution of the Kondo model amounts to
finding a special unitary operator, $U_S$, which
transforms the Kondo Hamiltonian $H_K(=H_f)$ into the 
``Toulouse limit Hamiltonian'' $H_T$,
\begin{equation} \label{US}
U^\dagger_S H^{}_K U^{}_S = H^{}_T \;,
\end{equation}
where $H_T$ is (in some sense) exactly solvable.
However, one may have neglected  irrelevant operators 
in performing the transformation (\ref{US}), see below.
The exactly solvable Toulouse limit Hamiltonian can in turn
be mapped onto a free fermion Hamiltonian by yet
another canonical transformation $U_T$,
\[
U^\dagger_T H^{}_T U^{}_T = H^{}_0\;,
\]
so that the canonical transformation $U$ relevant for the x-ray
edge problem is $U=U_S U_T$.
The $U_S$ transformation is known for the cases $k=2$ and
$k=4$. We shall briefly review these two 
derivations and refer the reader to the original
papers \cite{emery92,fg95} for a more detailed discussion.

{\em Two-channel Kondo problem.}
The relevant part of the bosonized Kondo Hamiltonian can
be written as 
\begin{eqnarray} \label{Ktwo}
H_K &=& H_0[\phi_s]+ H_0[\phi_{sf}] \\ \nonumber &+&
\frac{I_\perp}{2\pi a}\left[ s_+ e^{i\phi_s(0)}
\cos\phi_{sf}(0)+{\rm H.c.}\right]+
\frac{I_z}{2\pi}s_z\partial_x\phi_s(0)\;,
\end{eqnarray}
where the ``spin-flavor'' boson field is defined by
\[
\phi_{sf}(x)=\frac{1}{\sqrt{2}}\left[\phi_{1}(x)-
\phi_{2}(x)\right]\;.
\]
Emery and Kivelson \cite{emery92} noticed 
that after the unitary transformation $U_S= \exp[i s_z\phi_s(0)]$,
the Hamiltonian (\ref{Ktwo}) takes the form
\begin{equation} \label{TLtwo}
\tilde{H}_K= H_0[\phi_s]+ H_0[\phi_{sf}]+
\frac{I_\perp}{\pi a} s_x\cos\phi_{sf}(0)+
\frac{\lambda}{2\pi}s_z\partial_x\phi_s(0)
\end{equation}
with $\lambda=I_z-2\pi v_F$. Here $H_0$ denotes a 
free fermion Hamiltonian with Fermi velocity $v_F$.
The Toulouse limit corresponds to $\lambda=0$, leading to
$H_T=\tilde{H}_K|_{\lambda=0}$.
The $\lambda$ operator is irrelevant and therefore does
not contribute to the long-time asymptotics of the
$U_S$ correlation function.
Therefore Eq.~(\ref{TLtwo}) is essentially the Hamiltonian 
(\ref{xxxx}) for a $g=1/2$ Luttinger liquid containing an impurity. This
 can be seen by noticing that $s_x$
commutes with $H_T$ and hence can be set to $s_x=\pm 1/2$.
Alternatively, one can refermionize Eq.~(\ref{TLtwo}) to obtain
a Majorana resonant-level model.

{\em  Four-channel Kondo problem.}
In this case the representation (\ref{Jpsi}) for the spin currents
is not appropriate. A convenient representation which satisfies
the $SU(2)$ level-4 commutation relations (\ref{KM}) 
is \cite{fg95}
\begin{eqnarray} \label{Jfour}
J^-(x)&=&\frac{\sqrt{2}}{\pi a}\exp\left[\frac{i\phi_s(x)}{
\sqrt{2}}\right]\cos\left[\sqrt{\frac{3}{2}}
\phi_{sf}(x)
\right]\;, \\ \nonumber 
J_z(x)&=&\frac{1}{\sqrt{2}\,\pi} \partial_x\phi_s(x)\;.
\end{eqnarray}
Using the representation (\ref{Jfour}) in the Kondo
Hamiltonian and employing the transformation
$U_S=\exp[i s_z \phi_s(0) /\sqrt{2}]$,
one finds \cite{fg95}
\begin{eqnarray}
\label{TLfour}
\tilde{H}_T &=& H_0[\phi_s]+ H_0[\phi_{sf}] \\ &+&
\frac{I_\perp}{\pi a} s_x\cos\left[\sqrt{\frac{3}{2}}
\phi_{sf}(0)\right]+
\frac{\lambda}{2\pi} s_z\partial_x\phi_s(0)
\nonumber
\end{eqnarray}
with $\lambda=\sqrt{2}(I_z-\pi v_F)$.
The $\lambda$ operator is again irrelevant, and the Toulouse limit
Hamiltonian $H_T=\tilde{H}_K|_{\lambda=0}$
is exactly the Hamiltonian describing a $g=3/4$  Luttinger
liquid with an impurity, see Eq.~(\ref{xxxx}).

Therefore, the canonical transformation involved in the 
Toulouse limit solution is always a ``spin rotation''
$U_S= \exp[i \kappa s_z \phi_s(0)]$
with $\kappa=1$ for $k=2$ and $\kappa=1/\sqrt{2}$ for $k=4$.
The rotation angle is determined by the total spin field 
$\phi_s(0)$.
Provided possible irrelevant operators can be neglected,
the total spin field commutes with the impurity part of the 
Toulouse limit Hamiltonians.
Consequently, the correlation function of $U_S$
must be computed with respect to the ground state 
of $H_0[\phi_s]$ which coincides with
the ground state of $H_T$ as far as
the field $\phi_s$ is concerned.
Noticing that it does not matter in which spin state
$s_x=\pm \frac12$ the $U_S$ correlation function is computed,
and that the transformation $U_T$ is 
related to the spin-flavor field $\phi_{sf}$
(which commutes with $\phi_s$),
we obtain the factorized form at long times,
\begin{equation} \label{Dfac}
D(t)= \langle i| U^\dagger(t)U(0)|i \rangle =
\langle U^\dagger_S (t)U^{}_S(0) \rangle 
\langle U^\dagger_T (t)U^{}_T(0) \rangle \;,
\end{equation}
where $\langle U^\dagger_S (t)U^{}_S(0) \rangle \sim t^{-\kappa^2/4}$.
The second factor in Eq.~(\ref{Dfac})
is easily obtained since the correlation
function of $U_T$ determines the orthogonality
exponent in a Luttinger liquid.\cite{foot}
Therefore, we obtain for arbitrary $k$ the 
correlator
$\langle U^\dagger_T(t) U^{}_T(0)
\rangle \sim t^{-1/8}$.

In the end, we arrive at the following expression for the orthogonality
exponent of the Kondo problem,
\begin{equation}
\alpha_k=\frac{\kappa^2}{4}+\frac{1}{8}\;.
\end{equation}
This equation establishes the relationship between the 
orthogonality exponents of the multi-channel Kondo
impurity problem and the Luttinger liquid. 
In the two-channel case we have
\[
\alpha_{k=2}=\frac{1}{4}+\frac{1}{8}=\frac{3}{8} \;,
\]
while for the four-channel case
\[
\alpha_{k=4}=\frac{1}{8}+\frac{1}{8}=\frac{1}{4} \;.
\]
This is in agreement with Eq.~(\ref{alphaK}) obtained in 
Ref.\onlinecite{affleck94}.
Unfortunately, so far Toulouse-limit type solutions have not been
found for other values of $k$.

\section{Conclusions} \label{concl}

In this work, we have carried out an exact calculation
of the Fermi-edge singularity exponent in a Luttinger liquid.
The result $\alpha=1/8$ determines both
the scaling of the overlap $\langle f| i \rangle\sim 
N^{-\alpha/2}$ of the wavefunctions with and without the local
perturbation, and the asymptotic long-time
behavior of the core-hole Green's function,
$D(t) \sim \exp[-i\Delta E\, t]\, t^{-\alpha}$.
Our result is in notable contrast to the calculation by
Oreg and Finkel'stein \cite{oreg96} but verifies
 the result obtained previously from the open boundary 
fixed point assumption.\cite{gogolin93,kane94,prokofev94,affleck94}
The latter assumption therefore holds both for static and transient local
perturbations. 
In addition, we have extended the previously found relations 
between the multi-channel Kondo problem and the
problem of an impurity in the Luttinger liquid 
to the case when these impurities are dynamic.

{\em Note:} After completion of this manuscript, we became aware of 
the simultaneous and independent
work of Furusaki.\cite{furusaki97} Using essentially
the same method, he has also studied the orthogonality
exponent in a $g=1/2$ Luttinger liquid and obtained the same result.

\acknowledgements
We would like to acknowledge useful discussions with 
A.~A.~Nersesyan and A.~M.~Tsvelik. 
This work has been supported by the EPSRC of the U.K.~and
by the Deutsche Forschungsgemeinschaft (Bonn).


\begin{references}
\bibitem{anderson67} P.W. Anderson, Phys. Rev. Lett. {\bf 18}, 1049 (1967).
\bibitem{nozieres69} P. Nozieres and C.T. De Dominicis, Phys. Rev. {\bf 178},
1097 (1969).
\bibitem{schotte69} K.D. Schotte and U. Schotte, Phys. Rev.
{\bf 182}, 479 (1969); K. Sch\"onhammer, Z. Phys. B {\bf 45}, 23 (1981).
\bibitem{mahan90} G.D. Mahan, {\em Many-Particle Physics}
 (Plenum, New York, 1990); 
K. Ohtaka and Y. Tanabe, Rev. Mod. Phys. {\bf 62}, 929 (1990).
\bibitem{hewson93} A.C. Hewson, {\em The Kondo Problem to Heavy
Fermions} (Cambridge University Press, 1993).
\bibitem{luttinger} J.M. Luttinger, J. Math. Phys. {\bf 4}, 1154 (1963);
D.C. Mattis and E.H. Lieb, {\em ibid.} {\bf 6}, 304 (1965). 
\bibitem{haldane81} F.D.M. Haldane, J. Phys. C {\bf 14}, 2585 (1981).
\bibitem{schulz95} H.J. Schulz, in {\em Mesoscopic Quantum Physics},
Les Houches Session LXI, ed. by E. Akkermans {\em et al.} (Elsevier 1995).
\bibitem{exp} J.M. Calleja {\em et al.}, Solid State Commun.
{\bf 79}, 911 (1991);
M. Fritze {\em et al.}, Surface Science {\bf 305}, 580 (1994).
\bibitem{tarucha95} S. Tarucha, T. Honda, and M. Saku,
Solid State Commun. {\bf 94}, 413 (1995).
\bibitem{yacoby96} A. Yacoby {\em et al.},
 Phys. Rev. Lett. {\bf 77}, 4612 (1996).
\bibitem{lee92} D.K.K. Lee and Y. Chen, Phys. Rev. Lett. {\bf 69}, 1399 (1992).
\bibitem{ogawa92} T. Ogawa, A. Furusaki, and N. Nagaosa, Phys. Rev. Lett.
{\bf 68}, 3638 (1992).
\bibitem{gogolin93} A.O. Gogolin, Phys. Rev. Lett. {\bf 71}, 2995 (1993).
\bibitem{kane94} C.L. Kane, K.A. Matveev, and L.I. Glazman,
Phys. Rev. B {\bf 49}, 2253 (1994).
\bibitem{prokofev94} N.V. Prokof'ev, Phys. Rev. B {\bf 49}, 2148 (1994).
\bibitem{affleck94} I. Affleck and A.W.W. Ludwig, J. Phys. A
 {\bf 27}, 5375 (1994). 
\bibitem{oreg96} Y. Oreg and A.M. Finkel'stein, Phys. Rev. B {\bf 53},
10 928 (1996).
\bibitem{qin96} S. Qin, M. Fabrizio, and Lu Yu, Phys. Rev. B 
{\bf 54}, R9643 (1996).
\bibitem{fabrizio95} C.L. Kane and M.P.A. Fisher,
Phys. Rev. B {\bf 46}, 15 233 (1992); 
M. Fabrizio and A.O. Gogolin, {\em ibid.}
{\bf 51}, 17 827 (1995).
\bibitem{guinea85} F. Guinea, Phys. Rev. B {\bf 32}, 7518 (1985).
\bibitem{weiss88} U. Weiss and M. Wollensak, Phys. Rev. B 
{\bf 37}, 2729 (1988); U. Weiss, R. Egger, and M. Sassetti,
{\em ibid.} {\bf 52}, 16 707 (1995).
\bibitem{matveev95} K.A. Matveev, Phys. Rev. B {\bf 51}, 1743 (1995).
See also A. Furusaki and K.A. Matveev, {\em ibid.} {\bf 52}, 
16 676 (1995).
\bibitem{saleur96} A. Leclair, F. Lesage, and H. Saleur,
Phys. Rev. B {\bf 54}, 13 597 (1996).
\bibitem{toulouse69}
Originally, the Toulouse limit was discovered for the one-channel model in 
G. Toulouse, C. R. Acad. Sci. {\bf 268}, 1200 (1969).
\bibitem{emery92} V.J. Emery and S. Kivelson, Phys. Rev. B
{\bf 46}, 10 812 (1992).
\bibitem{fg95} M. Fabrizio and A.O. Gogolin, Phys. Rev. B
{\bf 50}, 17 732 (1995).
\bibitem{mus53} I. Gohberg and N. Krupnik, {\em One-Dimensional
Linear Singular Integral Equations} (Birkh\"auser, Basel, 1992);
N.I. Muskhelishvili, {\em Singular Integral Equations} (Nordhoff,
Groningen, 1953).
\bibitem{hamann71} D.R. Hamann, Phys. Rev. Lett. {\bf 26}, 1030 (1971).
\bibitem{comment} The Hamiltonian (\ref{HKfin}) 
only involves the spin (exchange) scattering; the charge (potential)
scattering is decoupled and can be treated in a usual 
way.\cite{nozieres69,mahan90}
\bibitem{foot} In fact, the operator $U_T$,  or rather the
universal part of this operator, has been 
constructed in Ref.\onlinecite{prokofev94}.
\bibitem{furusaki97} A. Furusaki, preprint (cond-mat/9702195).
\end{references}
\end{document}